\documentclass[aps,preprint,nofootinbib]{revtex4}
\begin{document}

\title{On the Schr\"{o}dinger equation for the free gravitational field} 

\author{Alexei M. Frolov}
\email{afrolov@uwo.ca}

\affiliation{Department of Applied Mathematics, University of Western Ontario, N6A 5B7, London,
             Canada }


\begin{abstract}
The Hamiltonian of the metric General Relativity derived in our earlier study (Gravitation 
{\bf 17}, 314 - 323 (2011)) is analyzed by the methods of Matrix Quantum Mechanics. This 
Hamiltonian is a quadratic function  of the momenta $\pi^{mn}$ conjugate to the spatial 
components $g_{mn}$ of the metric tensor $g_{\alpha\beta}$. The Hamiltonian is reduced to 
the new form which is more convenient to derive the Schr\"{o}dinger equation for the free 
gravitational field. In turn, this Schr\"{o}dinger equation is used to study possible 
motions of the free gravitational field(s). In particular, it is shown that harmonic 
oscillations of the free gravitational field, or harmonic gravitational waves, cannot 
be observed as an actual motion of this field. We also investigate the internal structure 
of the solitary gravitational wave. 
\end{abstract}


\date{\today}

\maketitle

\newpage

\section{Introduction}

In our earlier study \cite{Our1} we derived the following total Hamiltonian $H_T$ of the free gravitational 
field
\begin{equation}
  H_{T} = H_{c} + g_{00,0} \phi^{00} + 2 g_{0k,0} \phi^{0k} \; \; \; , \; \; \label{eq0}
\end{equation}
where $H_c$ is the `canonical', or `dynamical' part of the total Hamiltonian (or dynamical Hamiltonian, for 
short), $\phi^{00}$ and $\phi^{0k}$ are the primary constraints \cite{Dirac1}, while $g_{00,0}$ and $g_{0k,0}$ 
are the corresponding velocities, i.e. time derivatives (or temporal derivatives) of the corresponding 
components of the metric tensor $g_{\alpha\beta}$ \cite{Carm}. The dynamical Hamiltonian $H_c$ in Eq.(\ref{eq0}) 
takes the form
\begin{eqnarray}
 H_{c} &=& \frac{1}{\sqrt{-g} g^{00}} I_{mnpq} \pi^{mn} \pi^{pq} - \frac{1}{g^{00}} 
 I_{mnpq} \pi^{mn} B^{( p q 0 \mid \mu \nu k )} g_{\mu\nu,k} \nonumber \\
 &+& \frac14 \sqrt{-g} \Bigl[  \frac{1}{g^{00}} I_{mnpq} B^{((m n)  0 \mid \mu \nu k )} 
 B^{( p q 0 \mid \alpha \beta l )} - B^{\mu \nu k \alpha \beta l} \Bigr]  g_{\mu\nu,k} 
 g_{\alpha\beta,l} \; \; \; , \; \; \label{eq1}
\end{eqnarray}
where $g_{\mu\nu}$ and $g^{\alpha\beta}$ are the covariant and contravariant components of the metric tensor 
\cite{Carm}, respectively. Here and everywhere below in this study the Latin alphabet is used for spatial 
components and index ``0'' for a temporal component. The notations $\pi^{mn}$ in Eq.(\ref{eq1}) stand for 
momenta conjugate to the spatial components $g_{mn}$ of the metric tensor \cite{Our1}. The definition of 
these `spatial' $\pi^{mn}$ (where $m \ge 1$ and $n \ge 1$) and `temporal' momenta $\pi^{m0}$ (or $\pi^{n0}$) 
and $\pi^{00}$ can be found in \cite{Our1}. In general, all such momenta are designated as $\pi^{\alpha\beta}$, 
while the notation $g_{\rho\nu}$ stands for the components of the metric tensor. 

The notation $I_{mnpq}$ used in Eq.(\ref{eq1}) is
\begin{eqnarray}
    I_{mnpq} = \frac{1}{d-2} g_{mn} g_{pq} - g_{mp} g_{nq} = I_{pqmn} \; \; \; , \label{I}
\end{eqnarray}
where $d$ ($d \ge 2$) is the total dimension of the space-time continuum. The $I_{mnpq}$ quantity is the 
spatial (symmetric) tensor which is the $(d - 1)-$tensor (or $(d-1) \otimes (d-1)$-tensor) in the 
$d-$dimensional space-time. These values are different from the actual (or complete) $d-$tensors. The 
$d-$tensor $B^{\alpha\beta\gamma \mu\nu\rho}$ in Eq.(\ref{eq1}) is written in the following form \cite{Our1}
\begin{eqnarray}
  B^{\alpha\beta\gamma \mu\nu\rho} = g^{\alpha\beta} g^{\gamma\rho} g^{\mu\nu} 
 - g^{\alpha\mu} g^{\beta\nu} g^{\gamma\rho} + 2 g^{\alpha\rho} g^{\beta\nu} g^{\gamma\mu}  
 - 2 g^{\alpha\beta} g^{\gamma\mu} g^{\nu\rho} \label{B}
\end{eqnarray}
The symmetrized tensor $B^{(\alpha\beta\gamma \mid \mu\nu\rho)}$ is defined as a half of the
following sum:
\begin{eqnarray}
 B^{(\alpha \beta \gamma \mid \mu \nu \rho)} = \frac12 ( B^{\alpha \beta \gamma \mu \nu \rho} 
 + B^{\alpha \beta \rho \mu \nu \gamma} ) \; \; \; , \; \; \label{BS}
\end{eqnarray}
while the tensor $B^{((\alpha \beta) \gamma \mid \mu \nu \rho)}$ from Eq.(\ref{eq1}) is the 
symmetrized sum of the following two/four terms
\begin{eqnarray}
 B^{((\alpha \beta) \gamma \mid \mu \nu \rho)} = \frac12 \Bigl[ B^{(\alpha \beta \gamma \mid \mu 
 \nu \rho)} + B^{(\beta \alpha \gamma \mid \mu \nu \rho)} \Bigr]
 = \frac14 ( B^{\alpha \beta \gamma \mu \nu \rho} 
 + B^{\alpha \beta \rho \mu \nu \gamma} + B^{\beta \alpha \gamma \mu \nu \rho} 
 +  B^{\beta \alpha \rho \mu \nu \gamma} ) \; \; \; . \; \;  \label{BS1}
\end{eqnarray}

All $\frac{d (d + 1)}{2}$ components of the metric tensor $g_{\alpha\beta}$ and momenta $\pi^{\alpha\beta}$
conjugate to these components are the $d (d + 1)$ dynamical (and canonical) variables in the Hamiltonian 
approach. The fundamental (classic) Poisson brackets between the $g_{\alpha\beta}$ components of metric 
tensor and $\pi^{\alpha\beta}$ components of the `tensor of momentum' are \cite{Myths}
\begin{eqnarray}
 \Bigl\{ g_{\alpha\beta}, \pi^{\mu\nu} \Bigr\} = \frac12 ( \delta^{\mu}_{\alpha} 
 \delta^{\nu}_{\beta} + \delta^{\nu}_{\alpha} \delta^{\mu}_{\beta} ) = 
 \Delta^{\mu\nu}_{\alpha\beta} = - \Bigl\{\pi^{\mu\nu}, g_{\alpha\beta} 
 \Bigr\} \; \; \; , \; \label{Pois}
\end{eqnarray}
while the remaining Poisson brackets between these dynamical variables ($\pi^{\mu\nu}$ and 
$g_{\alpha\beta}$) are
\begin{eqnarray}
 \Bigl\{ g_{\alpha\beta}, g_{\mu\nu} \Bigr\} = 0 \; \; \; and \; \; \; 
 \Bigl\{ \pi^{\alpha\beta}, \pi^{\mu\nu} \Bigr\} = 0 \; \; \; . \; \;  \label{brack2}
\end{eqnarray}
In general, the knowledge of Poisson brackets between all dynamical variables in Hamiltonian system allows one 
to determine the actual `trajectories' of the field(s), i.e. all components of the metric tensor 
$g_{\alpha\beta}(t, \overline{x})$ as time-dependent functions given in each spatial point $\overline{x}$. 

In this study the properties of the $d (d + 1)-$canonical (Hamiltonian) variables $\Bigl\{ g_{00}, \ldots, 
g_{\alpha\beta}, \ldots, \pi^{00}, \ldots, \pi^{\mu\nu}, \ldots \Bigr\}$, Eqs.(\ref{Pois}) - (\ref{brack2}), are 
used to reduce the dynamical $H_c$ and total $H_T$ Hamiltonians to the forms which are more appropriate for the 
following analysis and derivation of the Schr\"{o}dinger equation for the free gravitational field. It also appears 
that the explicit form of the total Hamiltonian $H_T$ written in these canonical variables allows one to investigate 
the propagation of gravitational perturbations, or gravitational waves. In particular, it is shown below that each 
elementary gravitational perturbation propagates as a solitary wave with a steep front. It looks similar to 
the propagation of a strong, non-linear thermal wave. However, an important difference between a solitary
gravitational wave and strong thermal wave follows from the two facts: (a) all components of the metric tensor 
$g_{\alpha\beta}$ can change their values only at the front of the propagating gravitational wave, and (b) in the 
area behind the front these components are essentially constants. It follows from here that the components of the
metric tensor will never change again unless the next gravitational wave will change them. In other words, for 
propagating gravitational waves we do not face the problem of exhaustion of the source. It also is interesting to 
note that harmonic oscillations (i.e. vibrations) of the gravitational field itself play no role in its propagation 
(see below).

\section{Transformations of the dynamical Hamiltonian}

The dynamical Hamiltonian $H_c$ given by Eq.(\ref{eq1}) is a quadratic function of the spatial 
components of momenta $\pi^{mn}$. Let us transform this Hamiltonian $H_c$ to a slightly different
form which is more appropriate for the purposes of this study. First, note that Eq.(\ref{eq1}) can 
be re-written in the form
\begin{eqnarray}
 \tilde{H}_c &=& \sqrt{-g} g^{00} H_{c} = I_{mnpq} \pi^{mn} \pi^{pq} - \sqrt{-g}
 I_{mnpq} B^{( p q 0 \mid \mu \nu k )} g_{\mu\nu,k} \pi^{mn} \nonumber \\
 &+& \frac14 (-g) \Bigl[  \frac{1}{g^{00}} I_{mnpq} B^{((m n)  0 \mid \mu \nu k )} 
 B^{( p q 0 \mid \alpha \beta l )} - g^{00} B^{\mu \nu k \alpha \beta l} \Bigr] g_{\mu\nu,k} 
 g_{\alpha\beta,l} \label{eq2} \\
 &-&  \sqrt{-g} I_{mnpq} \Bigl\{\pi^{mn}, B^{( p q 0 \mid \mu \nu k )} g_{\mu\nu,k} \Bigr\}
 \nonumber
\end{eqnarray}
where the last term is the Poisson bracket of the $\pi^{mn}$ momenta and the product of the $B^{( p q 0 \mid 
\mu \nu k )}$ and $g_{\mu\nu,k}$ values. This bracket can be computed analytically with the use of Eq.(\ref{Pois}) 
and explicit formula for the $B^{( p q 0 \mid \mu \nu k )}$ quantity, Eq.(\ref{BS}). This leads to the result
\begin{eqnarray}
 \Bigl\{\pi^{mn}, B^{( p q 0 \mid \mu \nu k )} g_{\mu\nu,k} \Bigr\} &=& \frac12 \Bigl\{\pi^{mn}, 
 B^{p q 0 \mu \nu k} \Bigr\} g_{\mu\nu,k} + \frac12 \Bigl\{\pi^{mn}, B^{p q k \mu \nu 0} \Bigr\} 
 g_{\mu\nu,k} \nonumber \\ 
 &+& B^{( p q 0 \mid \mu \nu k )} \Bigl\{\pi^{mn}, g_{\mu\nu,k} \Bigr\} \label{eqPois}
\end{eqnarray}
where the Poisson brackets in the first and second terms of this equation are determined by Eq.(\ref{B}). This reduces 
each of the Poisson brackets from Eq.(\ref{eqPois}) to the sum of the following brackets
\begin{eqnarray}
  \Bigl\{ \pi^{mn}, B^{\alpha \beta \gamma \mu \nu \rho} \Bigr\} &=& 
  \Bigl\{ \pi^{mn}, g^{\alpha\beta} g^{\gamma\rho} g^{\mu\nu} \Bigr\}
  -  \Bigl\{ \pi^{mn}, g^{\alpha\mu} g^{\beta\nu} g^{\gamma\rho} \Bigr\} 
  + 2 \Bigl\{ \pi^{mn}, g^{\alpha\rho} g^{\beta\nu} g^{\gamma\mu} \Bigr\} \nonumber \\
  &-& 2 \Bigl\{ \pi^{mn}, g^{\alpha\beta} g^{\gamma\mu} g^{\nu\rho} \Bigr\} \label{eq11}
\end{eqnarray}
Note that all terms in the right-hand side of this equation have identical structure. Therefore, to illustrate our
calculations we can consider just one of these terms. For instance, for the first term in Eq.(\ref{eq11}) one finds
\begin{eqnarray}
  \Bigl\{ \pi^{mn}, g^{\alpha\beta} g^{\gamma\rho} g^{\mu\nu} \Bigr\} = 
   \Bigl\{ \pi^{mn}, g^{\alpha\beta} \Bigr\} g^{\gamma\rho} g^{\mu\nu} + 
   g^{\alpha\beta} \Bigl\{ \pi^{mn}, g^{\gamma\rho} \Bigr\} g^{\mu\nu} +
   g^{\alpha\beta} g^{\gamma\rho} \Bigl\{ \pi^{mn}, g^{\mu\nu} \Bigr\} 
\end{eqnarray}
where the Poisson bracket between the momentum $\pi^{mn}$ and $g^{\alpha\beta}$ component of the metric tensor
is calculated with the use of the following formulas
\begin{eqnarray}
  \Bigl\{ \pi^{mn}, g^{\alpha\beta} \Bigr\} =  \Bigl\{ \pi^{mn}, g^{\alpha}_{\alpha^{\prime}} 
 g^{\beta}_{\beta^{\prime}} g_{\alpha^{\prime}\beta^{\prime}} \Bigr\}  
 = - g^{\alpha}_{\alpha^{\prime}} g^{\beta}_{\beta^{\prime}} \Delta^{\mu\nu}_{\alpha^{\prime}\beta^{\prime}}
 = -\Delta^{\mu\nu;\alpha\beta}  
\end{eqnarray}
where the last equality contains the definition of the new delta-function (or delta-tensor) $\Delta^{\mu\nu;\alpha\beta}$ 
which contains the four upper indexes only. 

The Poisson bracket in the third term from Eq.(\ref{eqPois}) is 
\begin{eqnarray}
   \Bigl\{\pi^{mn}, g_{\mu\nu,k} \Bigr\} = - \Bigl(\Delta^{\mu\nu}_{\alpha\beta}\Bigr)_{,k}
\end{eqnarray}
where the $\Delta^{\mu\nu}_{\alpha\beta}$-function (or $\delta-$tensor) is defined above (see, Eq.(\ref{Pois})). 
Note that this Poisson bracket equals zero identically, if the delta-tensor $\Delta^{\mu\nu}_{\alpha\beta}$ is
considered as a constant term. This correspond to the classical approach. In quantum approach this term can be 
transformed to the form where the derivative upon spatial components appears in the front of the wave function.
Then, by integrating by parts we can move the spatial derivative from the delta-function to the wave function
$\Psi$. This means that such a term is not equal zero identically. By calculating all Poisson brackets in 
Eq.(\ref{eqPois}) one finds that the explicit formula for the Poisson bracket in the last term of Eq.(\ref{eq2}) 
contains a very large number of terms. It is not convenient in actual calculations. To avoid this problem below
we shall keep the united notation for the Poisson bracket in the last term of Eq.(\ref{eq2}). 

Now we note that in the Hamiltonian $H_c$, Eq.(\ref{eq2}), all momenta $\pi^{mn}$ are located at the very right 
position in each term. Such form of $H_c$ has a number of advantages to perform quantization of the 
classical system with the Hamiltinian $\tilde{H}_c$ given by Eq.(\ref{eq2}). The Hamiltonian $\tilde{H}_c$ can be 
represented as the product of the two spatial tensors (or $(d-1)-$tensors) $I_{mnpq}$ and $\tilde{H}^{pqmn}_{c}$, 
where the spatial tensor $I_{mnpq}$ is defined in Eq.(\ref{I}) and spatial tensor $\tilde{H}^{pqmn}_{c}$ is
\begin{eqnarray}
 \tilde{H}^{pqmn}_{c} &=& \pi^{pq} \pi^{mn} - \sqrt{-g} 
  B^{( p q 0 \mid \mu \nu k )} g_{\mu\nu,k} \pi^{mn} \nonumber \\
 &-& \frac{g}{4} \Bigl[ B^{((m n)  0 \mid \mu \nu k )} 
 B^{( p q 0 \mid \alpha \beta l )} - g^{00} E^{pqmn} B^{\mu \nu k \alpha \beta l} \Bigr] g_{\mu\nu,k} 
 g_{\alpha\beta,l} \label{eq4} \\
 &-& \sqrt{-g} \Bigl\{\pi^{pq}, B^{( m n 0 \mid \mu \nu k )} g_{\mu\nu,k} \Bigr\} \nonumber
\end{eqnarray}
where the spatial tensor $E^{pqmn}$ included in this equation is defined by the relation $I_{mnpq} E^{pqkl} = 
\delta^{k}_{m} \delta^{l}_{n}$ (or $\hat{I} \hat{E} = 1$). Components of this spatial tensor are the spatial 
components of the complete $E^{\mu\nu\gamma\sigma}$-tensor which is defined by the following equation (see, 
e.g., \cite{Our1})
\begin{equation}
   E^{\mu\nu\gamma\sigma} = e^{\mu\nu} e^{\gamma\sigma} - e^{\mu\gamma} e^{\nu\sigma} = E^{\gamma\sigma\mu\nu}
\end{equation} 
where 
\begin{equation}
   e^{\mu\nu} = g^{\mu\nu} - \frac{g^{0\mu} g^{0\nu}}{g^{00}} = e^{\nu\mu}
\end{equation} 
and, therefore:
\begin{equation}
   E^{\mu\nu\gamma\sigma} =  g^{\mu\nu} g^{\gamma\sigma} - g^{\mu\gamma} g^{\nu\sigma}
 - \frac{1}{g^{00}} \Bigl( g^{0\mu} g^{0\nu} g^{\gamma\sigma} + g^{\mu\nu} g^{0\gamma} g^{0\sigma}
 - g^{\mu\gamma} g^{0\nu} g^{0\sigma} - g^{0\mu} g^{0\gamma} g^{\nu\sigma} \Bigr)
\end{equation} 

Let us assume that we have performed the quantization of the Hamiltonian, Eq.(\ref{eq4}). In detail the process of
quantization is discussed in the next Section, but here we just want to make a few important comments about the explicit 
form of the Schr\"{o}dinger equation. As follows from our analysis above the arising Schr\"{o}dinger equation is written 
in the following `matrix' form with the Hamiltonian (spatial tensor) $\tilde{H}^{pqmn}_{c}$ from Eq.(\ref{eq4})
\begin{eqnarray}
 \imath \hbar E^{pqmn} \frac{\partial \Psi}{\partial \tau} = \tilde{H}^{pqmn}_{c} \Psi \label{EqSh2}
\end{eqnarray}
where $E^{pqmn}$ is the spatial tensor defined above. Both these operators ($E^{pqmn}$ and $\tilde{H}^{pqmn}_{c}$) in
Eq.(\ref{EqSh2}) are $g-$dependent (or metric-dependent) spatial tensors. This means that each component of these spatial 
tensors is a function of the components of metric tensor $g_{\mu\nu}$. The time $\tau$ in Eq.(\ref{EqSh2}) is related with 
the incident time $t$ by the relation $\tau = \frac{t}{\sqrt{-g} g^{00}}$. The new Hamiltonian (spatial tensor!) 
$\tilde{H}^{pqmn}_{c}$ in Eq.(\ref{eq4}) is a quadratic function of the momenta $\pi^{pq}$ and $\pi^{mn}$. Note that all 
momenta $\pi^{pq}$ and $\pi^{mn}$ which are included in the Hamiltonian $\tilde{H}^{pqmn}_{c}$ do not have temporal 
components. In other words, the $\tilde{H}^{pqmn}_c$ Hamiltonian does not contain any of the $\pi^{00}, \pi^{0m}$ and/or
$\pi^{n0}$ momenta. Now we need to perform the last step of our procedure and transform the classical expression for the 
$(d-1)-$tensor (or spatial tensor) $\tilde{H}^{pqmn}_{c}$, Eq.(\ref{eq4}), into the corresponding quantum operator.

\section{Quantization}

The goal of this Section is the quantization of the classical Hamiltonian $\tilde{H}^{pqmn}_{c}$ from Eq.(\ref{eq4}). 
The main step in this process is to replace all classical momenta $\pi^{\alpha\beta}$ and `coordinates' $g_{\mu\nu}$
in the classical Hamiltonian $\tilde{H}_c$ by the corresponding quantum operators. The classical Poisson bracket 
between each pair of these dynamical variables $\Bigl\{g_{00}, \ldots, g_{\alpha\beta}, \ldots, \pi^{00}, \ldots, 
\pi^{\mu\nu}, \ldots \Bigr\}$ is also replaced by the corresponding quantum Poisson bracket which explicitly contains the 
reduced Planck constant $\hbar$. Unfortunately, in many cases such a formal replacement of classical quantities by the 
corresponding quantum operators may lead to ambiguous answers. To avoid a possible appearance of multiple answers and produce 
the correct quantum expression one needs to apply the `correspondence principle' known in Quantum Mechanics since the middle 
of 1920's (see, e.g., \cite{LLQ}). For the free gravitational filed the correspondence principle means that the quantum 
Poisson bracket must have the correct limit in the case of very weak gravitational fields, or, in other words, for the 
flat space-time. This determines the following expression for the quantum (Q) Poisson bracket
\begin{eqnarray}
 \Bigl\{ g_{\alpha\beta}, \pi^{\mu\nu} \Bigr\}_Q = \imath \hbar \Bigl\{ g_{\alpha\beta}, \pi^{\mu\nu} \Bigr\}_C
 = \imath \hbar \frac12 ( \delta^{\mu}_{\alpha} \delta^{\nu}_{\beta} + \delta^{\nu}_{\alpha} 
 \delta^{\mu}_{\beta} ) = \imath \hbar \Delta^{\mu\nu}_{\alpha\beta} \label{PoisQ}
\end{eqnarray}   
From here one can write the following explicit formula for the quantum operator of momentum $\pi^{\alpha\beta}$ in the 
$g_{\alpha\beta}$-representation (i.e. in the `coordinate' representation)
\begin{equation}
 \pi^{\mu\nu} = - \imath \hbar \Bigl[ \frac{\partial}{\partial g_{\mu\nu}} + f_{\mu\nu}(g_{\alpha\beta}) \Bigr]
 \label{moment}
\end{equation}
where $f_{\mu\nu}(g_{\alpha\beta})$ is a regular (or analytical) function of all components of the metric tensor. The quantum 
operators of momenta, Eq.(\ref{moment}), must also obey the basic relations given by Eq.(\ref{brack2}) which are true for both 
the classical and quantum Poisson brackets. This leads to a set of additional conditions for the $f_{\mu\nu}$-functions from 
Eq.(\ref{moment}) 
\begin{equation}
 \frac{\partial f_{\mu\nu} }{\partial g_{\alpha\beta} } = \frac{\partial f_{\alpha\beta} }{\partial g_{\mu\nu} }
 \label{cond1}
\end{equation}
In general, one can use some freedom to choose different types of the $f_{\mu\nu}$ functions in Eq.(\ref{moment}) 
to simplify either the definition of momenta $\pi^{\mu\nu}$, or the formula for the quantum Hamiltonian operator 
$\tilde{H}^{pqmn}_{c}$ in Eq.(\ref{EqSh2}). In reality, such a freedom is quite restricted, since there are a number 
of rules for canonical transformations which can only be used to transform one set of dynamical (Hamiltonian) variables into 
another. This is true for arbitrary Hamiltonian systems, including systems with constraints (for more details, see \cite{Our1} 
and \cite{Myths}). To avoid discussion of this problem, which is not directly related with our goals in this study, below 
we shall assume that the additional function $f_{\mu\nu}(g_{\alpha\beta})$ in Eq.(\ref{moment}) equals zero identically. 
It can be shown that such a choice is `natural' for Hamiltonian formulation of GR originally developed in \cite{Pirani} 
and later corrected in \cite{Our1}.  

Substitution of the operators of momenta $\pi^{\mu\nu} = \imath \hbar \frac{\partial }{\partial g_{\mu\nu} }$ in the 
classical Hamiltonian tensor $\tilde{H}^{pqmn}_{c}$, Eq.(\ref{eq4}), produces the quantum Hamiltonian operator 
$\hat{\tilde{H}}^{pqmn}_{c}$ which is correct at least in the lowest-order approximation upon $\hbar$ \cite{Dirac1}, 
\cite{Tut}. With this Hamiltonian operator we can write the following Schr\"{o}dinger equation \cite{Schrod} 
\begin{eqnarray}
 \imath \hbar E^{pqmn} \frac{\partial \Psi}{\partial \tau} &=& \hat{\tilde{H}}^{pqmn}_{c} \Psi = 
 - \hbar^2 \frac{\partial^2 \Psi}{\partial g_{pq} \partial 
 g_{mn}} - \imath \hbar \Bigl[\sqrt{-g} B^{( p q 0 \mid \mu \nu k )} g_{\mu\nu,k}\Bigr] 
 \frac{\partial \Psi}{\partial g_{mn}}  \nonumber \\
 &-& \frac{g}{4} \Bigl( \Bigl[ B^{( p q 0 \mid \alpha \beta l )} B^{((m n)  0 \mid \mu \nu k )} 
  - g^{00} E^{pqmn} B^{\mu \nu k \alpha \beta l} \Bigr]  g_{\mu\nu,k} 
 g_{\alpha\beta,l} \Bigr) \Psi \label{EqSh3} \\
 &-& \sqrt{-g} \Bigl\{\pi^{mn}, B^{( p q 0 \mid \mu \nu k )} g_{\mu\nu,k} \Bigr\} \Psi \nonumber
\end{eqnarray}
where, in general, the wave function $\Psi = \Psi(\tau, \{ g_{\alpha\beta} \})$ depends upon all components of the metric 
tensor $g_{\alpha\beta}$ and time $\tau$. This equation describes time-evolution of the free gravitational field which is 
now considered as a `quantum object' and described by the wave function $\Psi$. It should be mentioned that this equation 
is only one from a number of conditions (or equations) which must be obeyed for the actual wave function $\Psi$ of the free 
gravitational field(s). These additional conditions are the primary constraints and all other constraints which arise during 
time-evolution of the primary constraints \cite{Dirac1950}. Formally, such additional conditions for the wave function $\Psi$ 
follow from the fact that the Schr\"{o}dinger equation must contain the total Hamiltonian $H_T$, rather than the dynamical 
Hamiltonian $H_c$. This was well understood and emphasized by Dirac in 1950 \cite{Dirac1950}.   

Let us discuss these additional conditions for the wave function $\Psi$ of the free gravitational field. First, consider the 
conditions which follow from the primary constraints. As mentioned above the actual Schr\"{o}dinger equation must contain the 
total Hamiltonian $H_T$, Eq.(\ref{eq0}). Its replacement with the dynamical Hamiltonian $H_c$ is possible if (and only if), the 
following conditions are obeyed for the wave functions $\Psi$ 
\begin{equation}
    \phi^{00} \Psi = 0  \; \; \;  and \; \; \;  \phi^{0k} \Psi = 0 \label{primary}
\end{equation}
for $k = 1, 2, \ldots, d-1$, where $d$ is the total dimension of the space-time continuum. These conditions are the $d-$primary 
constraints written explicitly in \cite{Our1}. The primary constraints arise, since the corresponding componenet of moments $\pi^{00}$
and $\pi^{0k}$ cannot be defined from the original singular Lagrangian \cite{Our1}. By replacing the momenta and coordinates in these 
primary constraints by the corresponding quantum operators one finds
\begin{equation}
 \phi^{0\sigma} = \pi^{0\sigma} - \frac12 \sqrt{-g} B^{((p q) 0 \mid \mu \nu k)} g_{\mu\nu,k} = \imath \hbar 
  \frac{\partial }{\partial g_{0\sigma} } - \frac12 \sqrt{-g} B^{((p q) 0 \mid \mu \nu k)} g_{\mu\nu,k}
\end{equation}
Therefore, the $d-$primary constraints, Eq.(\ref{primary}), can be written in the form of the following differential equations 
for the wave function $\Psi$: 
\begin{equation}
 \imath \hbar \frac{\partial }{\partial g_{0\sigma} } \Psi = \frac12 \sqrt{-g} B^{((p q) 0 \mid \mu \nu k)} 
 g_{\mu\nu,k} \Psi \label{cond2}
\end{equation}
where $\sigma = 0, 1, \ldots, d - 1$. 

As was mentioned by Dirac \cite{Dirac1} the primary constraints are absolute, i.e. they must hold at arbitrary time. This 
means that the equations which govern the time evolution of the primary constraints must be time independent. For quantum system 
with constraints this lead to the chain of the following equalities:
\begin{eqnarray}
 \chi^{0\sigma} \Psi &=&  \phi^{0\sigma}_{,0} \Psi = \{ \phi^{0\sigma}, H_T \} \Psi = 0 \; \; \; , \nonumber \\
 (\chi^{0\sigma})_{,0} \Psi &=& (\phi^{0\sigma}_{,0})_{,0} \Psi = \{  \{ \phi^{0\sigma}, H_T \}, H_T \} \Psi = 0  \; \; , \ldots 
 \label{eq27} 
\end{eqnarray}
For  actual physical fields the arising chains of related equalities, i.e. constraints, are always finite \cite{Dirac1}, \cite{Tut}. 
Furthermore, the values $\chi^{0\sigma} = \phi^{0\sigma}_{,0} = \{ \phi^{0\sigma}, H_T \}$ are called the secondary constraints 
\cite{Dirac1950}, while their Poisson brackets with the total Hamiltonian $H_T$ are the tertiary constraints, etc. Analysis of 
the constrained structure of the field equations in metric GR indicates clearly that the free gravitational field has $d-$primary 
constraints $\phi^{0\sigma}$ and $d-$secondary constraints $\chi^{0\sigma}$ for $\sigma = 0, 1, \ldots, d - 1$. The secondary 
constraints $\chi^{0\sigma}$ are independent of each other and of any of the primary constraints. It was shown in \cite{Our1} 
that the time-evolution of the secondary constraints $\chi^{0\sigma}$ lead to the linear combinations of the same secondary 
constraints. The coefficients in such linear combinations depend upon spatial components of the metric tensor, their spatial 
derivatives and primary constraints \cite{Our1}. This fact proves the closure of the Dirac procedure for the free gravitational 
field in metric GR, since no tertiary constraints have been found. The closed analytical formulas for the secondary constraints 
$\chi^{0\sigma}$ have been found in a number of earlier papers (see, e.g., \cite{Our1} and references therein). These formulas 
are extremely complicated and here we do not want to repeat them, since they are not crucially important for our analysis below.

It should be mentioned that the Poisson brackets between the primary and secondary constraints of the free gravitational field(s) 
are \cite{Our1}
\begin{eqnarray}
 \Bigl\{ \chi^{0\sigma}, \phi^{0\gamma} \Bigr\} = - \frac12 g^{\gamma\sigma} \chi^{00} \label{prisec}
\end{eqnarray}
i.e. it is proportional to the secondary constraint $\chi^{00}$. Therefore, if $\Psi$ is the solution of the 
Schr\"{o}dinger equation, Eq.(\ref{EqSh3}), then for such wave functions one finds $\chi^{0\sigma} \phi^{0\gamma} \Psi = 
\phi^{0\gamma} \chi^{0\sigma} \Psi = 0$. In other words, the operators of the primary and secondary constraints commute with each
other on the solutions $\Psi$ of the Schr\"{o}dinger equation. 

Thus, for the wave function $\Psi$ of the free gravitational field in addition to the Schr\"{o}dinger equation, Eq.(\ref{EqSh3}),
we have $d-$primary and $d-$secondary constrints. In general, for $d-$dimensional space-time we have $\frac{d ( d + 1 )}{2}$ 
independent components of the metric tensor $g_{\alpha\beta}$. Therefore, the total number of freedoms $f$ of the free gravitational 
field in metric GR is $f = \frac{d ( d + 1 )}{2} - 2 d = \frac{d (d - 3)}{2}$. As follows from this formula the actual non-constraint 
motion becomes possible for the free gravitation field when $d \ge 4$. In particular, the free gravitational field in our universe 
($d = 4$) has two degrees of freedom. For $d = 3$ we have $f = 0$, i.e. only constrained motion can be found at such dimension of 
space-time. Note also that in addition to the primary and secondary constraints for many Hamiltonian systems one also finds a 
number of `conservation laws' which must be obeyed for any real motion. 

\section{Gravitational waves. Hamiltonian factorization.}

In reality it is very difficult to obtain any closed (analytical) solution of the Schr\"{o}dinger equation, Eqs.(\ref{EqSh3}), 
(\ref{primary}) and (\ref{eq27}) for the wave function $\Psi$. The complexity of this problem is absolutely outstanding. However, 
there is a group of problems in metric GR which can be investigated directly with the use of the total Hamiltonian $H_T$, or the 
corresponding quantum operator $\hat{H}_T$. These problems are closely related with the propagation of gravitational perturbations.  

In modern literature the propagation of any gravitational perturbation is always considered as a propagation of harmonic 
oscillations, or, in other words, harmonic waves. Moreover, it is emphasized explicitly that the propagation of the gravitational 
perturbations, or waves, is very similar to the propagation of the electromagnetic waves in vacuum. In reality, such gravitational 
oscillatory waves have never been detected in various experiments extensively conducted since the middle of 1950's. Nevertheless, a 
large number of people are still trying to observe such waves and measure their basic properties. Note that the first theoretical 
prediction of the oscillatory gravitational waves was made by Einstein in his paper published in 1918 \cite{Einst1918}. However, in 
\cite{Einst1918} Einstein considered the case of very weak gravitational fields. In this approximation the actual four-dimensional 
space-time was essentially replaced by the flat space-time. In \cite{Einst1918} Einstein wrote that his conclusion about 
propagating oscillatory gravitational waves is substantially based on the linear approximation used and it can be false in the 
actual metric GR. Later, Einstein and his co-workers arrived to a conclusion that the oscillating gravitational waves cannot 
exist in the metric GR, but their paper submitted in Phys. Rev. was rejected and manuscript was lost. Probably, some parts of the 
original text of that paper were included in \cite{Einst1937}. Other details related with Einstein's opinion about oscillating 
gravitational waves in the metric GR can be found on the Web-page \cite{Blog}.      

The Hamiltonian approach for metric GR developed in \cite{Our1} allows one to re-consider the problem of propagating gravitational 
waves. Let us assume that at some point of the Universe we have a gravitational process, e.g., collision of the two stars which lead 
to the formation of the new star. Gravitational fields around this collision area change rapidly. Briefly, we can say that in such a
case we deal with perturbations of the gravitation field(s), or, in other words, with gravitational perturbations. As known from 
Astrophysics typical collisions of actual stars proceed dozens of years (and even longer) and gravitational waves are generated at each 
moment of this process. It is cleat that our Hamiltonians (see, e.g., Eq.(\ref{EqSh3})) contain only gravitational fields (or $g_{\alpha\beta}$ 
components) and we cannot describe, in principle, actual collisions of stars and/or any other process related with the finite-time 
redistribution of masses in space. To avoid an unwanted discussion of the phenomena which cannot be analyzed by our methods we have to 
define an elementary gravitational perturbation which is considered as an infinitesimal part of the real (i.e. finite) process of 
gravitational changes. Everywhere below by an `elementary gravitational perturbation' we mean the process of actual gravitation motion 
which is local and takes an infinitely small time $\delta t$. The corresponding, infinitesimal changes in the gravitational fields 
$g_{\mu\nu}$ can be described with our Hamiltonian approach. By using the language from the Differential Equations one can say that here we 
are trying to determine and investigate the corresponding Green's function(s). We can also introduce a closely related definition of the 
solitary gravitational wave as a wave which transfrers an elementary gravitational perturbation and produces changes in the gravitational 
fields, i.e. in the components of the metric tensor $g_{\alpha\beta}$. Our main goal here is to determine the laws which govern the propagation 
of the solitary gravitational wave in space-time, i.e. in the Universe. Also, we want to investigate the internal structure of such a 
wave. Thus, below we shall consider only elementary gravitational perturbations and solitary gravitational waves which move these 
perturbations from the point of their generation to the rest of the Universe. The actual gravitational perturbation can be represented 
as a superposition, i.e. sum and/or integral, of a large number (even infinite number) of elementary perturbations. The same statement 
is true for gravitational waves propagating from an actual source of gravity.  

Our analysis of the propagating solitary gravitational wave is based on the explicit form of the dynamical Hamiltonian $\tilde{H}_c$, 
Eq.(\ref{eq2}), or the corresponding Hamiltonian spatial tensor $\tilde{H}^{pqmn}_{c}$, Eq.(\ref{eq4}). Both these Hamiltonians are the 
quadratic expressions upon the momenta $\pi^{mn}$ conjugate to the spatial components $g_{mn}$ of the metric tensor $g_{\alpha\beta}$, 
i.e. the situation looks similar to the case of the free electromagnetic field propagating in vacuum (see, e.g., \cite{Heitl}). A 
complete similarity with electrodynamics will be observed, if (and only if) we can show that the Hamiltonian $\tilde{H}^{pqmn}_{c}$, 
Eq.(\ref{eq4}), is a quadratic function of the spatial components $g_{mn}$ of the metric tensor $g_{\alpha\beta}$. In this case by 
applying some standard methods from Matrix Quantum Mechanics (see, e.g., \cite{Green}) one can reduce this Hamiltonian to the sum of 
`quadratic' operators which is essentially coincides with the Hamiltonian of the harmonic oscillator and/or with the Hamiltonian free 
electromagnetic field which is used in Quantum Electrodynamics (see, e.g., \cite{Dirac1}, \cite{LLQE}, \cite{GelfFomin}). However, it can 
be shown (see below) that the dependence of the Hamiltonian tensor $\tilde{H}^{pqmn}_{c}$ in Eq.(\ref{eq4}) upon the components of the 
metric tensor $g_{\alpha\beta}$ is substantially more complex and cannot be represented as a quadratic function (i.e. quadratic polynomial) 
of the components of the metric tensor. 

To investigate this problem we write the Hamiltonian $\tilde{H}^{pqmn}_{c}$ in the form  
\begin{eqnarray}
   \tilde{H}^{pqmn}_{c} = \pi^{pq} \pi^{mn} + S^{pqmn} \label{quadr}
\end{eqnarray}
where the spatial tensor $S^{pqmn}$ takes the form 
\begin{eqnarray}
 S^{pqmn} &=& \imath \hbar \Bigl[\sqrt{-g} B^{( p q 0 \mid \mu \nu k )} g_{\mu\nu,k}\Bigr] 
 \frac{\partial}{\partial g_{mn}}  \nonumber \\
 &-& \frac{g}{4} \Bigl( \Bigl[ B^{( p q 0 \mid \alpha \beta l )} B^{((m n)  0 \mid \mu \nu k )} 
  - g^{00} E^{pqmn} B^{\mu \nu k \alpha \beta l} \Bigr]  g_{\mu\nu,k} 
 g_{\alpha\beta,l} \Bigr) \label{EqSh5} \\
 &-& \sqrt{-g} \Bigl\{ \pi^{mn}, B^{( p q 0 \mid \mu \nu k )} g_{\mu\nu,k} \Bigr\} \nonumber
\end{eqnarray} 
Now, we want to show explicitly that this spatial tensor $S^{pqmn}$ is not a quadratic function of the $g_{\alpha\beta}$ components 
(or $g_{mn}$ components) and/or it cannot be reduced to such a form. The formula for the $S^{pqmn}$ quantity, Eq.(\ref{EqSh5}), contains 
three different terms. First, consider the second term which is a polynomial function of the $g_{\alpha\beta}$ components. 
The maximal power of such a polynomial upon $g_{\alpha\beta}$ is $10 = 4 + 3 + 3$ (not 2), where four is the power of the 
determinant $g$ of the metric tensor $g_{\alpha\beta}$, while the power of each factor $B$ in this formula equals three. This (second) 
term in Eq.(\ref{EqSh5}) also contains the product of the two spatial derivatives $g_{\mu\nu,k} g_{\alpha\beta,l}$. The first and third 
terms in Eq.(\ref{EqSh5}), contain the factor $\sqrt{-g}$ which is, in fact, an algebraic function (not a finite polynomial!) of the 
$g_{\alpha\beta}$ components. In general, this functions is represented as a sum of the infinite number of powers of the $g_{\alpha\beta}$ 
components. This indicates clearly that we cannot reduce neither the Hamiltonian  $\tilde{H}^{pqmn}_{c}$ from Eq.(\ref{eq4}), nor the 
dynamical Hamiltonian $\tilde{H}_c$ from Eq.(\ref{eq2}) to a quadratic form which is needed for similarity with the Hamiltonian of the 
free electromagnetic field. By using the regular transformations of the metric tensor we can try to reduce the total power of the 
Hamiltonian $\tilde{H}^{pqmn}_{c}$ in Eq.(\ref{eq4}) upon $g_{\alpha\beta}$ components. However, it appears that such a power upon the 
spatial components of the metric tensor $g_{mn}$ always exceeds five. This result is of fundamental importance for the metric General 
Relativity, since it indicates clearly that the free gravitational fields cannot propagate in space-time as `harmonic vibrations' (or 
oscillations). 

In the case of very weak gravitational fields one can find a similarity with the free electromagnetic field. Indeed, for very weak 
gravitational fields the differences between the corresponding components of the metric tensor and Minkovskii tensor are small and 
$\sqrt{-g} = 1$. This allows to determine the limiting forms of these two Hamiltonians ($\tilde{H}^{pqmn}_{c}$ from Eq.(\ref{eq4}) and 
$\tilde{H}_c$ from Eq.(\ref{eq2})). The correct Hamiltonian transition to the case of the weak gravitational fields is described in 
\cite{Linear}. It appears that now both these Hamiltonians are quadratic functions of the new variables $h_{\alpha\beta}$, where 
$h_{\alpha\beta} = g_{\alpha\beta} - \eta_{\alpha\beta}$ are the small corrections to the corresponding components of the Minkowskii 
tensor $\eta_{\alpha\beta} = diag(-, +, +, \ldots, +)$ in the flat space-time. Formally, this means that for very weak gravitational 
fields there is a similarity with the electromagnetic fields. This includes similarity in propagation of the free gravitational waves and 
electromagnetic waves.  

For arbitrary gravitational fields we have $\sqrt{-g} \ne 1$ and the values $h_{\alpha\beta} = g_{\alpha\beta} - \eta_{\alpha\beta}$ are
not small. Briefly, this means that there is no close analogy with the electromagnetic fields. In particular, the propagation of gravity in 
space-time continuum has a number of fundamental differences with the propagation of electromagnetic radiation. It is clear that harmonic 
vibrations of the free gravitational filed(s) do not play any role in the propagation of the free gravitational fields. Moreover, such 
harmonic oscillations of the gravitational field do not exist as actual motions. This is the main result of our analysis. It appears 
that the propagation of the gravitational fields is described by the same Schr\"{o}dinger equation, Eq.(\ref{EqSh3}), with the Hamiltonian 
$\hat{\tilde{H}}^{pqmn}_{c}$ (spatial tensor) written explicitly in the right-hand side of this equation. By investigating this Hamiltonian
and comparing it with the well known Hamiltonian of the electromagnetic field (see, e.g., \cite{Dirac1} and \cite{Heitl}) one finds a few 
similarities and a number of fundamental differences. An obvious similarity is the quadratic dependence of each of these Hamiltonians upon
momenta $\pi^{mn}$ conjugate to the corresponding spatial components of the gravitational and electromagnetic field(s). The main fundamental
difference between these two Hamiltonians follows from the fact that the Hamiltonian in metric GR is a substantially non-linear function of 
the components of the metric tensor $g_{\alpha\beta}$, or field components, while the Hamiltonian of the free electromagnetic field is a 
simple quadratic function of the corresponding field components. Such differences in the two Hamiltonians lead to fundamentally different 
equations of motions for the free gravitational and free electromagnetic fields. In the last case the total Hamiltonian is represented as an 
infinite sum of one-dimensional Hamiltonians of harmonic oscillators. This is well known Fourier resolution of the free electromagnetic field 
(see, e.g., \cite{Heitl}, \cite{LLE}).  

\section{Propagation of the free gravitational fields}

As we have shown above harmonic vibrations of the free gravitational filed(s) do not represent actual motions of the free 
gravitational fields. This means that the propagation of the free gravitational field(s) cannot be represented as the propagation of the 
harmonic waves, or harmonic oscillations. This fundamental fact is of great importance for the future theoretical development of metric GR. 
For instance, it is clear now that all quantization procedures developed earlier for the free gravitational field(s) have no connection with 
reality, if they based on the of propagating harmonic waves, or, in other words, on systems of harmonic oscillators. However, from our  
astrophysical experience we know that regular gravitational fields cannot be bounded in one spatial area and they always propagate through 
the whole Universe. Therefore, the propagation of gravitational waves in the metric GR is real. Moreover, we can define the propagating
gravitational wave which can be represented as a decomposition of solitary waves. The goal of this Section is to analyze such solitary 
gravitational waves, their internal structure and propagation.

Here we want to discuss the actual propagation of the free gravitational fields by using the Hamiltonian approach described above. It appears 
that the laws which govern the propagation of the free gravitational field can be obtained from the classic total Hamiltonian $H_T$, 
Eq.(\ref{eq0}), and/or from the corresponding Schr\"{o}dinger equation with the quantum operator $\hat{H}_T$ which correspond to this total 
Hamiltonian $H_T$. To simplify our analysis here we restrict ourselves to the classical approach only. In \cite{Our1} the following formula 
for the classical total Hamiltonian $H_T$ has been derived
\begin{eqnarray}
 H_{T} = - 2 g_{0\sigma} \chi^{0\sigma} + g_{00,0} \phi^{00} + 2 g_{0k,0} \phi^{0k} + \Bigl[ 2g_{0m} \phi^{mk} 
  - \sqrt{-g} E^{mnki} g_{mn,i} + \nonumber \\
 \sqrt{-g} g_{\mu\nu,i} \frac{g^{0\mu}}{g^{00}} 
 \Bigl( g^{\nu k} g^{0i} - g^{\nu i} g^{0k} \Bigr) \Bigr]_{,k} \label{Hamiltt}
\end{eqnarray}
As follows from this equation, the total Hamiltonian $H_{T}$ is the sum of the terms proportional to the primary ($\phi^{00}$ and 
$\phi^{0k}$) and secondary ($\chi^{0\sigma}$) constraints. It is also contains the total spatial derivatives which are combined in 
one `surface' term. This surface term can be represented in a slightly different form with the use of the following spatial vector (or 
$(d-1)-$vector) $\overline{G} = (G^1, G^2, \ldots, G^d)$, where 
\begin{eqnarray}
 G^{k} = 2g_{0m} \phi^{mk} - \sqrt{-g} E^{mnki} g_{mn,i} + \sqrt{-g} g_{\mu\nu,i} \frac{g^{0\mu}}{g^{00}} 
 \Bigl( g^{\nu k} g^{0i} - g^{\nu i} g^{0k} \Bigr) \label{vector}
\end{eqnarray}
is the $k-$contravariant component of this $(d - 1)-$vector. The vector $\overline{G}$ is the energy flux of the free gravitational 
field, i.e. it determines the flow of the gravitational energy (or, gravitational flow, for short) through the closed boundary 
$(d-1)-$surface of the volume occupied by the gravitational field. Indeed, by taking the integral from both sides of Eq.(\ref{Hamiltt}) 
over the whole volume $V$ occupied by the gravitational and enclosed by the closed surface $S$ one finds   
\begin{eqnarray}
 E = \int div \overline{G} dV = - \oint (\overline{G} \cdot \overline{n}) dS_{d-1} = - \oint \overline{G} \cdot d\overline{S}_{d-1} 
 \label{gauss}
\end{eqnarray}
where $\overline{G}$ is the (d-1)-dimensional vector defined in Eq.(\ref{vector}), $\overline{n}$ is the unit vector of the outer normal to 
the surface element $dS_{d-1}$ and $d\overline{S}_{d-1} =  \overline{n} dS_{d-1}$ is the elementary volume of the surface $dS_{d-1}$ 
oriented in the direction of the outwardly directed normal $\overline{n} = (n_1, n_2, \ldots, n_d)$. To transform the integral in 
Eq.(\ref{gauss}) we have applied the Gauss formula for multi-dimensional integrals.

The gravitational vector $\overline{G}$, Eq.(\ref{vector}), plays the same role in metric General Relativity as the Pointing vector 
(or Umov-Pointing vector) plays in Electrodynamics \cite{LLE}. Note that the left-hand side of the energy conservation law in Electrodynamics 
contains the time-derivative of the total field energy, i.e. $\frac{\partial E}{\partial t}$, rather than the total field energy 
itself. The same general identity must be correct in the metric GR. To avoid possible contradictions we can transform the expression in 
the left-hand side of Eq.(\ref{gauss}) in the following way  
\begin{eqnarray}
 E &=& \int \frac{\partial E}{\partial t} \delta(t_f - t) dt = E_f = \frac{v_f}{c} \int \oint \Bigl(\frac{\partial w}{\partial t}\Bigr) 
 \delta(t_f - t) c dt dS_{d-1} \label{gauss1} \\
 &=& x c \int \oint \Bigl(\frac{\partial w}{\partial t}\Bigr) \delta(t_f - t) dt dS_{d-1} \nonumber
\end{eqnarray}
where $E_f$ is the energy at the front of the propagating gravitational wave, $w$ is the spatial density of the energy $E$, i.e. the energy 
per unit volume, i.e. $w = \lim_{V \rightarrow 0} \Bigl( \frac{E}{V} \Bigr)$, while $c$ is the speed of light in vacuum and $v_f$ is the 
propagation velocity of the gravitational wave in vacuum. Note that from Eq.(\ref{gauss1}) we have $E = E_f$, i.e. all energy of the 
propagating gravitational wave is concentrated in its front. Also, the time $t_f$ in this formula coincides with the time when the propagating 
gravitational wave will reach the boundary surface $S_{d-1}$ and the factor $x = \frac{v_f}{c}$. Very likely, that the velocity of the front 
propagation $v_f$ equals to the speed of light in vacuum exactly, i.e. $v_f = c$ and, therefore, the factor $x \Bigl(= \frac{v_f}{c}\Bigr)$ in 
Eq.(\ref{gauss1}) equals unity. However, such an assumption must be confirmed in a number of independent experiments. 

Thus, we have shown that all energy of the propagating gravitational wave is associated only with the front of such a wave. Before and after 
the wave front area the local gravitational enery, i.e. energy spatial density, is a constant (or zero). This conclusion follows from the fact
that the total Hamiltonian is zero before the wave front and it equals to the sum of constraints after the wave front. The only non-zero term in 
the total Hamiltonian $H_T$ describes the gravitational flow through the surface which was reached by the propagating gravitational wave. The
concentration of the whole energy of the propagating gravitational wave in its front is the direct consequence of the substantial non-linearity
of the field equations in metric GR. In some sense the propagating gravitational wave is very similar to the strong shock wave which propagates 
in a compressible gas mixture. This is a brief description of the internal structure of the propagating (solitary) gravitational wave. Such a 
structure is very simple, but it is clear that only this structure agrees with the original ideas of GR proposed and developed by Mach and 
Einstein. 

\section{Conclusions}

We have considered the Hamiltonian formulation of the metric General Relativity (or metric GR, for short) for the free gravitational field. By 
using the process of quantization and explicit form of the classical Hamiltonian \cite{Our1} we have derived the Schr\"{o}dinger equation for the 
free gravitation filed. The explicit forms of this Schr\"{o}dinger equation and $2 d-$additional differential conditions, which correspond to 
the $d-$primary and $d-$secondary constraints, for the wave function $\Psi$ are extremely complex to obtain any closed analytical solution. It 
is clear that this problem needs an additional investigation.  

Nevertheless, the Hamiltonian approach used in this study allows one to make some general predictions about propagation of the free gravitational 
fields. First, it is clear that gravitational waves considered as harmonic vibrations of the field itself do not exist in the metric GR (or 
Einstein GR). However, in the approximation of the weak gravitational filed (so-called linearized gravity) one finds motions of the field which 
can be considered as `vibrations' or reasonable approximations to vibrations of the components of some small tensors. It should be mentioned that 
this result was derived by Einstein in 1918 \cite{Einst1918} for very weak gravitational field(s), but all such harmonic `vibrations', or 
`oscillations' disappear quickly and completely, as soon as `linearized' GR is replaced by the real, i.e. non-linear, GR. An approximate criterion 
of such a non-linearity can be written in the form $\sqrt{- g} = 1$. If we consider the cases when $\sqrt{- g}$ cannot be replaced by unity, then 
harmonic oscillations of the gravitational field play no role in the propagation of this field. In particular, this includes the actual metric GR, 
where the gravitational fields are not very weak.

Second, the propagation of the free gravitational field(s) in space-time continuum is substantially different from propagation of the 
light wave in vacuum. The same Hamiltonian approach being applied to the free gravitational fields leads to the following conclusions about 
propagation of such fields: (1) gravitational fields created by one elementary (sudden and local) gravity perturbation at 
the origin begin to propagate to the rest of the Universe as a solitary gravitational wave with a steep front surface; (2) all energy and 
momentum of this gravitational wave are concentrated in the front of this wave and transferred by this front; (3) behind the front of 
the solitary gravitational wave we have a steady-state gravitational field distribution which corresponds to the final state of the 
gravitational source; (4) there is neither energy, nor momentum transfer behind the front of the propagating gravitational wave. Very likely, 
that any solitary gravitational wave is obtained from infinite number of harmonic oscillations during their propagation in the medium which is
substantially non-linear upon the spatial components of the metric tensor $g_{\alpha\beta}$. The arising internal structure of the solitary 
gravitational wave is the direct consequence of very substantial non-linearity of the original field equations.
 
Here we need to make a few following remarks. First, in this study we have analyzed only solitary gravitational waves created by elementary  
gravity perturbations. Furthermore, it was assumed that the gravitational wave arising during such a perturbation propagates into an empty 
space-time continuum. In this case the speed of such a gravitational wave must be equal to the speed of light in vacuum, however, such an assumption 
must be confirmed by a number of independent experiments. Otherwise, we need to introduce another fundamental velocity different from $c$. In respect 
to the fundamental ideas of metric GR proposed and developed by Mach and Einstein before the front of the first solitary gravitational wave 
propagating in vacuum we have no actual space-time. Such an spatial area can be considered as a continuum free from any gravitational fields. Behind 
the front of the propagating gravitational wave we have a steady-state gravitational filed(s) distribution which corresponds to the final 
distribution of gravity sources. 

The main result of this study is the analytical, Hamiltonian-based description of the free gravitational field which is free from internal 
contradictions. Dynamical variables in this approach are the components of the metric tensor $g_{\alpha\beta}$ (`coordinates') and momenta
$\pi^{\mu\nu}$ conjugate to them. Such a choice of the variables makes our dynamical system `natural' which means that the Lagrangian of this
system is a quadratic functions upon velocities, while its Hamiltonian is also a quadratic function of the space-like momenta $\pi^{mn}$. Based on 
the Hamiltonian approach we derived the Schr\"{o}dinger equation and analyzed the structure of a solitary gravitational which propagates in vacuum. 
The front of such a wave is an infinitely thin surface which moves away (with some finite speed) from the original perturbation area. At the front 
of this wave the components of the metric tensor $g_{\alpha\beta}$ are suddenly changed to their final values. The propagating solitary gravitational 
wave has only one advanced front and no reverse front et all (in contrast with the shock waves known from gasodynamics). This means that any 
elementary gravitational perturbation changes all components of the metric tensor $g_{\alpha\beta}$ in each spatial point once and forever. These 
values can be changed again only by the next solitary gravitational wave. Such a structure the propagating gravitational perturbation truly reflects 
the original ideas of the General Relativity proposed and developed by Mach and Einstein. It is also clear that the idea of oscillating gravitational 
waves contradicts these ideas. 

\begin{center}
   {\bf Appendix}
\end{center}

The goal of this Appendix is to show that the internal structure of the propagating gravitational wave described above is natural, i.e. it corresponds
to the nature of the propagating gravitational wave. First, we need to use the result of Einstein \cite{Einst1918} that very weak gravitational waves
essentially coinside with the harmonic oscillation waves, i.e. waves which are emitted by a set of harmonic oscillators. Second, the variables which
are used to describe such harmonic oscillations ($h_{\alpha\beta}$) are simply (linearly) related with the components of the metric tensor 
$g_{\alpha\beta}$ (see above). The third fundamental fact is a substantial non-linear dependence of the `gravitational velocities' upon the 
gravitational fields themselves. This means that the time-derivatives (or temporal derivatives) of the components of the metric tensor $g_{\alpha\beta}$ 
are the non-linear functions of these components. 

To prove the last statement let us consider the Lagrangian which is the $\Gamma-\Gamma$ part the Einstein-Hilbert (EH) Lagrangian. This Lagrangian is a
quadratic function of the first-order derivatives of the metric tensor (for more details see, e.g., \cite{Carm}) 
\begin{equation}
L = \sqrt{-g} g^{\alpha\beta} \Bigl( \Gamma_{\alpha\nu}^{\mu} \Gamma_{\beta\mu}^{\nu} - 
 \Gamma_{\alpha\beta}^{\nu} \Gamma_{\nu\mu}^{\mu} \Bigr) = \frac14 \sqrt{-g} B^{\alpha\beta\gamma\mu\nu\rho} g_{\alpha\beta,\gamma} 
 g_{\mu \nu,\rho} \label{Aeq1}
\end{equation}
where the coefficients $B^{\alpha\beta\gamma\mu\nu\rho}$ are defined in the main text. The same Lagrangian is re-written in terms of velocities, i.e. 
it is reduced to the form which explicitly contains the time derivatives of the metric tensor \cite{Our1}
\begin{eqnarray}
 L = \frac14 \sqrt{-g} B^{\alpha\beta0\mu\nu0} g_{\alpha\beta,0} g_{\mu\nu,0} + \frac12 \sqrt{-g} B^{\left( \alpha\beta0\mid\mu\nu k\right) }
 g_{\alpha\beta,0}g_{\mu\nu,k} + \frac14 \sqrt{-g} B^{\alpha\beta k\mu\nu l} g_{\alpha\beta,k} g_{\mu\nu,l} \; \; \; , \label{Aeq2}
\end{eqnarray}

Now, we can define the momenta $\pi^{\gamma\sigma}$ conjugate to the metric tensor $g_{\gamma\sigma}$
\begin{equation}
 \pi^{\gamma\sigma} = \frac{\delta L}{\delta g_{\gamma\sigma,0}} = \frac12 \sqrt{-g} B^{\left( \left( \gamma\sigma\right) 0\mid\mu\nu0\right) }
 g_{\mu\nu,0} + \frac12 \sqrt{-g} B^{\left( \left( \gamma\sigma\right)0\mid\mu\nu k\right) } g_{\mu\nu,k} \; \; \; . \label{Aeq3}
\end{equation}
It can be shown that if the both indexes $\gamma$ and $\sigma$ are space-like (i.e. $\gamma = m$ and $\gamma = n$), then Eq.(\ref{Aeq3}) is invertible 
and one finds the following expression for the velocity $g_{mn,0}$
\begin{equation}
 g_{mn,0} = I_{mnpq} \frac{1}{g^{00}} \Bigl( \frac{2}{\sqrt{-g}} \pi^{pq} - B^{((pq) 0\mid\mu\nu k)} g_{\mu\nu,k} \Bigr) \label{Aeq4}
\end{equation}
For non-singular dynamical systems we can always write $\pi^{pq} \approx G^{pqab} g_{ab,0}$. Therefore, from Eq.(\ref{Aeq4}) and explicit formulas for the 
$B^{((pq) 0\mid\mu\nu k)}$ coefficients, Eq.(\ref{B}), one finds that the velocity $g_{mn,0}$ is the non-linear function of the components of metric tensor 
$g_{\alpha\beta}$. This meas that all velocities of the space-like components of the metric tensor are the non-linear functions of $g_{\alpha\beta}$. Finally,
by combining all three facts mentioned above we arive to a uniform conclusion about the internal structure of the propagating gravitational wave
(see the main text). Note also, that our approach used in this study is based on the method originally developed in \cite{Pirani} and later corrected in 
\cite{Our1}. The approach developed by Dirac in \cite{Dirac} produces essentially the same results. The quantization procedure is even simpler in the Dirac
approach. In general, the equivalence of these two Hamiltonian procedures in metric GR was clear from the very beginning, since the dynamical variables in 
both these formulations are relted with each other by a canonical (Hamilton) transformation (for more details, see, \cite{Our1}, \cite{Myths}).  

\section{Acknowledgments}

I am grateful to my friends D.G.C. (Gerry) McKeon, N. Kiriyushcheva and S.V. Kuzmin (all from the University of Western Ontario, London, Ontario, CANADA) for 
helpful discussions and inspiration.

\end{document}